\documentclass[12pt,preprint]{aastex}

\shorttitle{M31 LBV }
\shortauthors{Humphreys et al. }

\begin{document}

\title{A New Luminous Blue Variable in M31\altaffilmark{1} }

\author{
Roberta M.. Humphreys\altaffilmark{2}, 
John C. Martin\altaffilmark{3}
and 
Michael Gordon\altaffilmark{2},  
}

\altaffiltext{1}  
{Based  on observations  with the Multiple Mirror Telescope, a joint facility of
the Smithsonian Institution and the University of Arizona.
} 

\altaffiltext{2}
{Minnesota Institute for Astrophysics, 116 Church St SE, University of Minnesota
, Minneapolis, MN 55455; roberta@umn.edu} 

\altaffiltext{3}
{Barber Observatory, University of Illinois, Springfield, IL, 62703}

\begin{abstract}
We report the fifth confirmed Luminous Blue Variable/S Doradus variable in M31.
In 2006, J004526.62+415006.3 had the spectrum of hot Fe II emission line star with 
strong P Cygni profiles in the Balmer lines \citep{Massey07}. In 2010, its absorption line
spectrum resembled an early A-type supergiant with H and Fe II emission lines with strong P Cygni profiles, and in 2013 the spectrum had fully transitioned to an F-type supergiant due to the 
formation
of the optically thick, cool wind which characterizes LBVs at maximum light. The photometric record supports the LBV/S Dor  nature of the variability. Its bolometric luminosity $\sim$ -9.65 mag places it 
on the HR Diagram near the known LBVs, AE And, Var C in M33 and S Dor. 
\end{abstract} 


\section{Introduction}
The class of stars known in the literature as Luminous Blue Variables(LBVs) are an important phase in the final evolution of evolved massive stars.
They may represent a short high mass losing stage in which the star sheds much of its outer envelope prior to becoming a Wolf-Rayet star, or as some authors 
have suggested, they may be the immediate progenitors of some types of 
Type II supernovae.   A  classical LBV or S Dor variable  in 
quiescence (minimum  light) resembles a moderately evolved hot star usually 
with the spectrum of a B-type supergiant or an Of/late WN star \citep{HD94,Vink}. During an LBV eruption, the mass loss rate increases, the wind becomes 
opaque and cool, and its absorption spectrum
resembles an A to F-type supergiant. Since this is a shift in the bolometric correction, the
star brightens in the visual and appears to move to the right, to lower temperatures on the
HR Diagram. This is the LBV's optically thick wind stage or maximum light.
The origin  of the instability that leads to this dramatic change is not 
understood, but most suggestions involve an instability close to the Eddington Limit. The mass loss event can last for several years or even decades, and
afterwards the star returns to it previous state.

Many stars are called LBVs or candidate LBVs, but  few are confirmed probably due to the infrequency of the LBV ``eruption'' or maximum light stage. 
For example, there are only four confirmed classical or normal LBV/S Dor variables in M31 (see \citet{RMH14}). Thus the discovery of a new LBV in its 
eruption state is important for understanding their relation to the larger 
massive star population, the frequency of the LBV eruption, and the duration 
of this unstable phase.  

J004526.62+415006.3\footnote{M31-004526.62 in \citet{RMH14} and \citet{Shol}}
was described by \citet{Massey07} as a ``hot LBV candidate" Its normalized spectrum from  2006 (Fig 12 in \citet{Massey07}) showed strong H emission lines with P Cygni profiles and weak Fe II emission lines. In a survey of luminous stars in M31 and M33, \citet{RMH14} noted that its 2010 spectrum closely resembled that of the warm hypergiant
J004444.52+412804.0 \citep{RMH13} with prominent P Cygni 
profiles in the multiplet 42 Fe II emission lines, strong hydrogen emission with broad 
wings and P Cygni profiles, and the absorption line spectrum  of an early 
A-type supergiant. A spectrum from 2011 published by  \citet{Shol} clearly showed that 
the Fe II lines had weakened. They also noted the change from 2006 and
suggested that it is an LBV. A second star J004051.59+403303.0 was identified
 as an LBV candidate.

In this paper we discuss recent spectra and photometry of 
J004526.62+415006.3  that confirm  it is an LBV currently in an optically 
thick wind state or maximum light.

\section{New Observations} 

In addition to the spectrum  described in \citet{RMH14}, it was observed again on 25 and 26  September  2013 with the Hectospec Multi-Object Spectrograph (MOS) \citep{Fab98,Fab05} on the 6.5-m MMT on Mt. Hopkins.  The Hectospec\footnote{http://www.cfa.harvard.edu/mmti/hectospec.html} has a 1$\arcdeg$ FOV and uses 300 fibers each with a core 
diameter of 250$\mu$m subtending 1$\farcs$5 on the sky. We used the 600 l/mm grating with the 
4800{\AA} tilt  yielding  $\approx$ 2500{\AA} coverage with 0.54 {\AA}/pixel resolution 
and R of $\sim$  2000. The same grating with a tilt of 6800{\AA} was used for the red 
spectra with $\approx$ 2500{\AA} coverage, 0.54{\AA}/pixel resolution and R of $\sim$ 3600.  
The spectra  were reduced using an exportable version of the CfA/SAO SPECROAD package 
for  Hectospec data\footnote{External SPECROAD was developed by Juan Cabanela for use
on Linux or MacOS X systems outside of CfA. It is available online at:
http://iparrizar.mnstate.edu.}. The spectra were all bias subtracted, flatfielded and 
 wavelength calibrated. 

Recent broadband CCD photometry was obtained by J. C. Martin with the
20-inch telescope at the Barber Observatory in 2013 and 2014, and by 
Sholukhova et al.(2014) in 2011. We also  
measured  additional magnitudes from   
 {\it HST}/ACS/HRC images observed August 2010 (HST-GO-12056) and January 2012 (HST-GO-12106).   The magnitudes  are zero pointed to Vega and comparable to Johnson B and Cousins I.

All of the spectroscopic and photometric observations discussed here are 
summarized in Table 1.

\section{The Spectrum  from 2006 to 2013 and Light Curve}

Phil Massey kindly provided his spectrum from 2006 \citep{Massey07} 
reproduced here in Figure 1. \citet{Massey07} described J004526.62+415006.3 as a
``hot LBV candidate". The spectrum is dominated by strong hydrogen emission
with broad wings and P Cygni profiles and Fe II and [Fe II] emission. Some weak absorption lines of He I $\lambda$4026, O II $\lambda$4070-4076 and S IV $\lambda$ 4116 are visible,  suggestive of an early B-type star. 

\begin{figure}[htb] 
\figurenum{1}
\epsscale{1.0}
\plotone{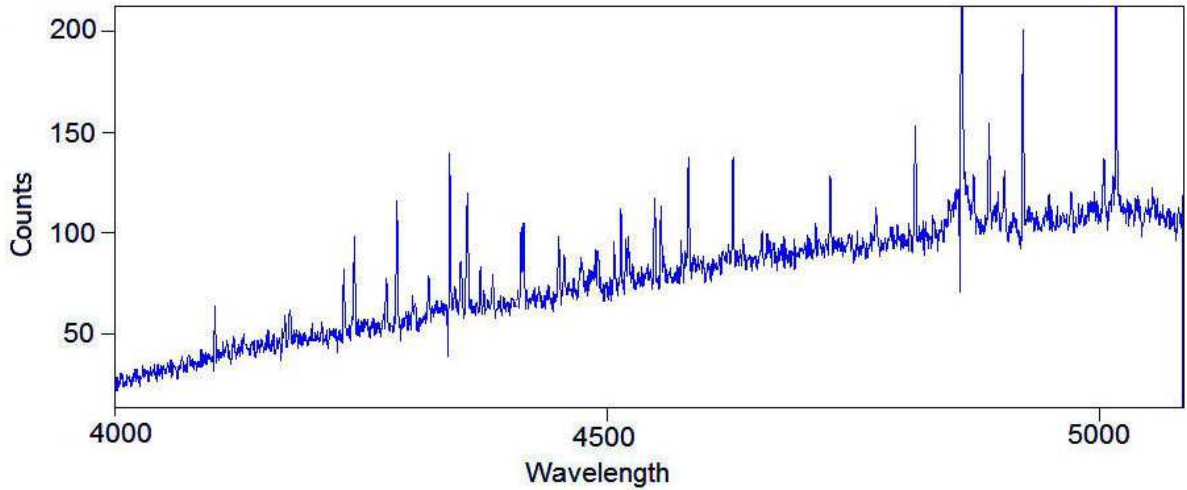}  
\caption{The Fe II and hydrogen emission line spectrum from 2006 
\citep{Massey07}.}
\end{figure}

The blue and red spectra from 2010 and 2013 are shown 
together in Figures 2a and 2b, respectively. They not only illustrate the dramatic change 
from the hot emission line star in 2006 but also the transition to a cooler temperature from 2010 to 2013 and the formation of the optically thick cool wind in
the 2013 spectrum.  \citet{RMH14} classified its 2010 spectrum  as an early 
A-type supergiant (A2eIa). With its strong P Cygni profiles in the multiplet 
42 Fe II emission lines they noted its close resemblance to the 
warm hypergiant J004444.52+412804.0 \citep{RMH13}.  

\begin{figure}[htb]
\figurenum{2a}
\epsscale{1.0}
\plotone{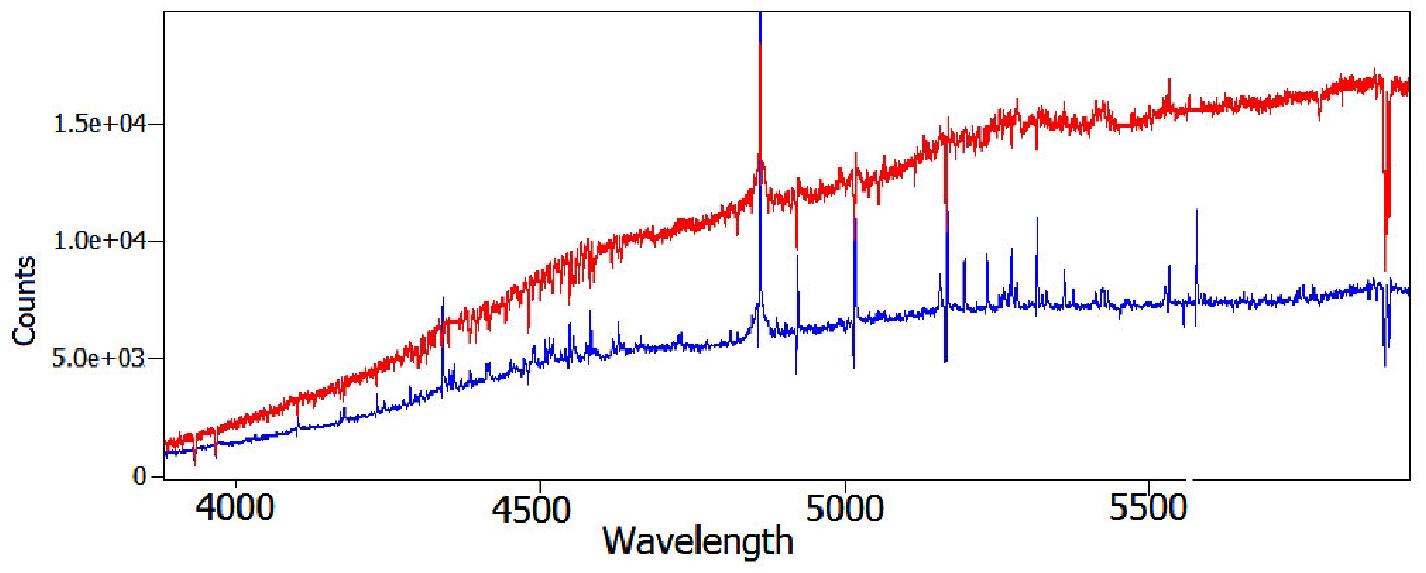} 
\caption{The blue spectra from 2010 (blue)  and 2013 (red). Note the 
development of the absorption line spectrum replacing the Fe II emission lines 
and weakening of the P Cygni profiles.}
\end{figure}

\begin{figure}[htb]
\figurenum{2b}
\epsscale{1.0}
\plotone{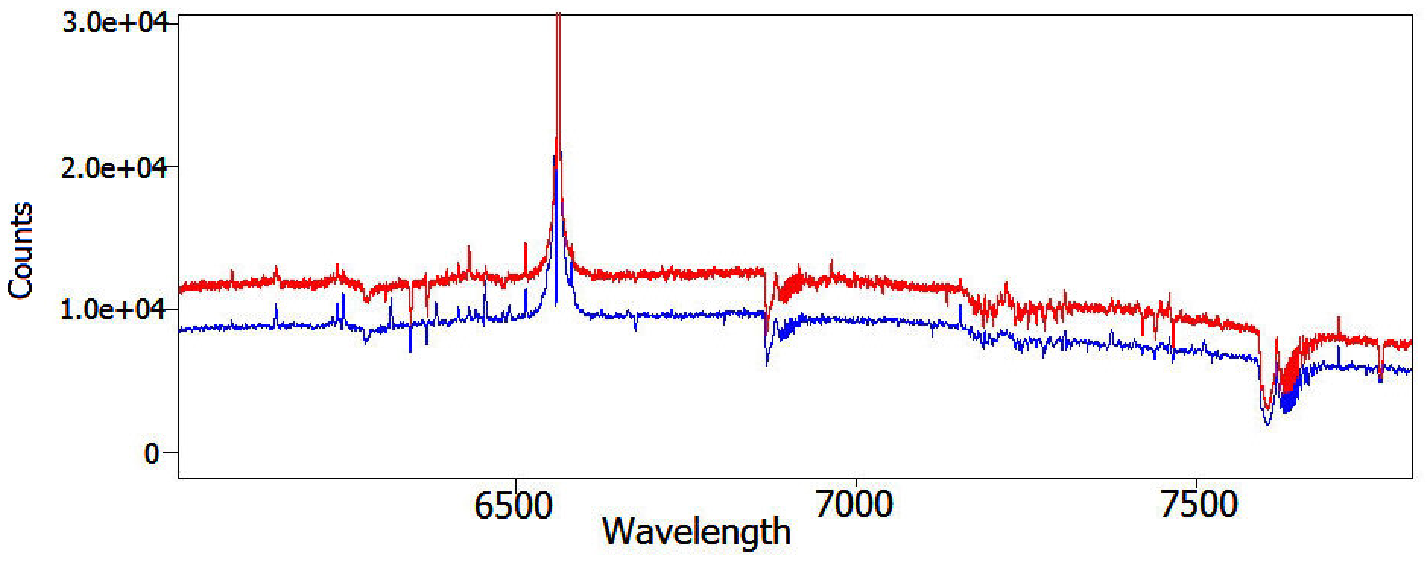}
\caption{Red spectra from 2010 (blue)  and 2013 (red). The luminosity sensitive O I triplet at $\lambda$7774{\AA} in intermediate temperature supergiants is 
much stronger in 2013.}
\end{figure}

The 2013 spectra show the  continued evolution of the wind to cooler temperatures. Metallic absorption lines have replaced the Fe II emission spectrum, strong Ca II H and K lines are conspicuous,
and the P Cygni profiles are gone 
except for H$\alpha$. The blue absorption line spectrum of the optically
thick wind is consistent with an early F-type supergiant ($\sim$ F2). In the red
spectrum,  the luminosity sensitive O I $\lambda$7774 triplet and the N I (multiplet 3) absorption lines are  much stronger. The differences between the 2010 and 2013 spectra illustrate 
the spectroscopic transition and the formation of an optically thick wind typical of LBV/S Dor variables.  

Like other LBVs \citep{RMH14,Oksala,Kraus}, its spectral energy distribution has a small near-infrared excess due to free-free emission, but it was not detected in the IRAC survey of M31 \citep{Mould}  or the WISE Source Catalog. So it  apparently does not have a near- or mid-infrared excess due to dust (see Figure 11 in \citet{RMH14}). However,  the relative strengths of the [N II]
$\lambda$5755{\AA} and $\lambda$6584{\AA} emission lines compared to the [S II] lines in the 2010 spectrum
indicate the presence
of a relatively dense circumstellar nebula, much denser than expected for an H II region \citep{Weis}.
The outflow velocities from the stellar wind measured from the absorption
minima in the P Cygni profiles in these spectra are summarized in Table 3. 
LBVs not only have slow wind velocities during their  maximum light, 
dense wind state, but during their quiescent hot state
their winds are also slower than those  measured the same way in normal supergiants of similar
spectral type \citep{RMH14}. J004526.62+415006.3  shows the same pattern, although
the outflow measured in the 2010 spectrum  is interestingly higher 
when the star was in transition to the dense wind state.

The light curve is shown in Figure 3.  Unfortunately, we do not
have either spectra or  photometry between 2006 and 2010 to determine 
the onset of the LBV/S Dor eruption. However, our spectra demonstrate that
the star was in transition to its cool dense wind state in 2010, so the 
outburst probably began in 2010 or shortly before that. 
Based on the photometry from  1999-2003  to 2012-2014, 
$\Delta$V is surprisingly, at most, only 0.9 mag which is too small for the implied
change in apparent temperature between the 2006 and 2010--2013 spectra. 
Thus we suspect that the earlier photometry is not representative of its minimum light or quiescent state, and  it may have been in an 
elevated state at that time.  Previous  photometry also shows
that it has had other eruptions; it was measured at V 
= 16.3 in 1992 \citep{Magnier}. 

\begin{figure}[htb] 
\figurenum{3}
\epsscale{1.0}
\plotone{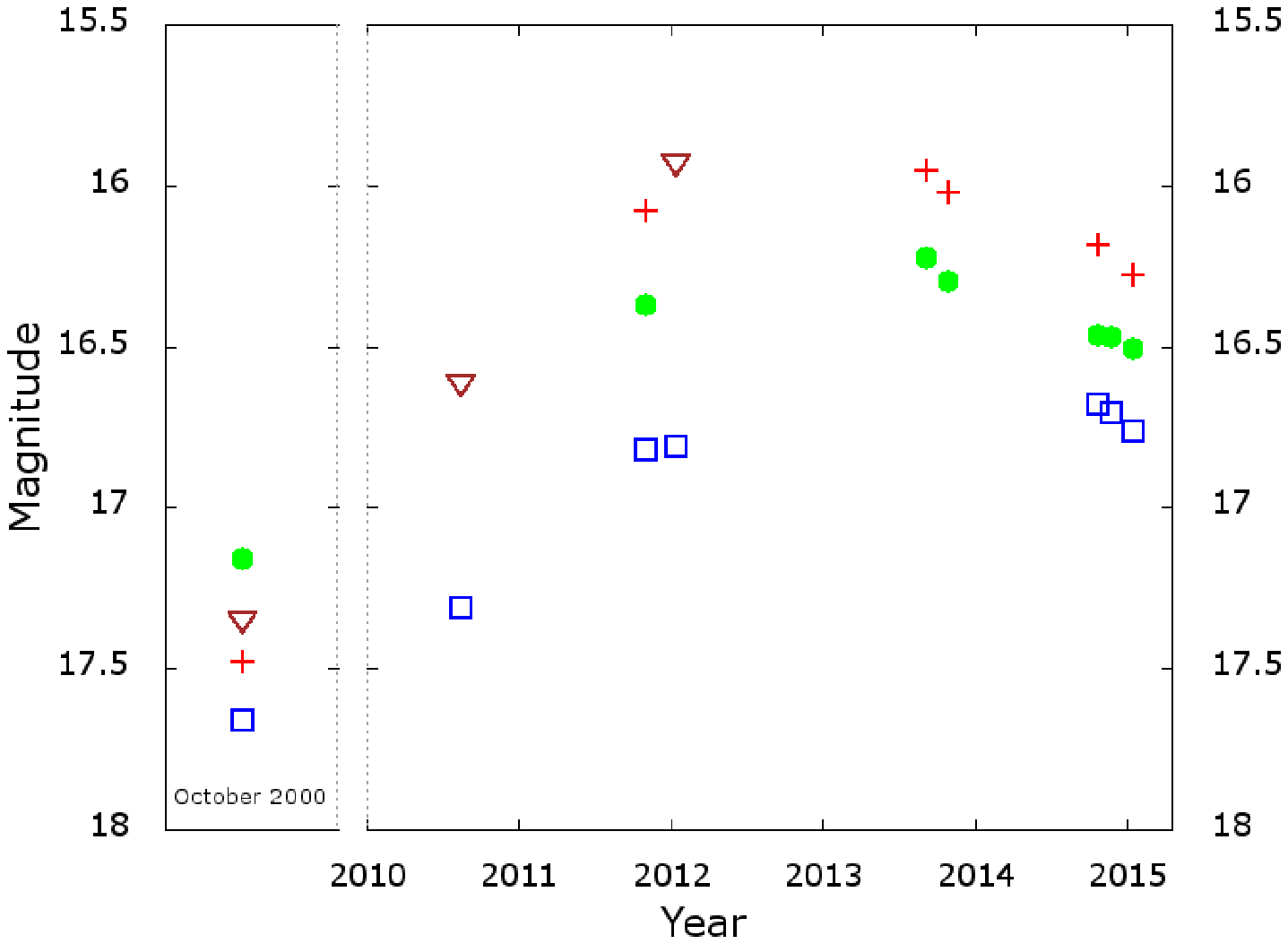}
\caption{The light curve based on the the data in Table 2. The symbols are B, 
open blue squares; V, filled green circles, R, red crosses and I, open red 
triangles. A gap is show n between 2000 and next available data in 2010.}
\end{figure}

\section{Relation to Other LBVs}

Adopting the interstellar extinction, A$_{v}$ = 1.5 mag 
from  \citet{RMH14}, measured from  three different indicators, its visual 
luminosity at maximum light is M$_{v}$ -9.65 with a distance modulus of 24.4 mag for M31 \citep{Riess}. At  maximum light, the bolometric correction will be quite small, corresponding to the early F-type  spectrum of the dense wind. 
 M$_{bol}$ is therefore also $\approx$ -9.65 mag. The corresponding parameters for the star in its quiescent or hot state are less well determined, but its spectrum from 2006 suggests an early B type supergiant with a surface temperature on the order of 20,000 K or hotter.  On the HR Diagram, J004526.62+415006.3 will thus lie very 
close to AE And in M31, Var C in M33, and the well studied progenitor of
the class S Dor, see Figure 12 in \citet{RMH14}. 

Given the lack of photometric data between 2000 and 2010, any additional data
from that period or earlier would be appreciated to better define the properties of the progenitor and the onset of the current eruption, as well as future photometric and spectroscopic monitoring.

We are especially grateful to Phil Massey for sharing his spectrum  of J004526.62+415006.3  from 2006 and to Olga Sholukova and her collaborators for a preprint of their paper.  We also think the Hectospec support operators Perry Berlind and Mike Calkins. Research by R. Humphreys on massive stars is supported by  
the National Science Foundation AST-1109394. J. C . Martin's collaborative work on luminous variables is supported by the National Science Foundation grant  AST-1108890.

{\it Facilities:} \facility{MMT/Hectospec, HST/ACS/HRC}

%%%%%%%%Tables

\begin{deluxetable}{llllll}
\rotate
\tablewidth{0 pt}
\tabletypesize{\footnotesize}
\tablenum{1} 
\tablecaption{Journal of Observations Used in this Paper}
\tablehead{
\colhead{UT Date} &
\colhead{Instrument}  &
\colhead{Exp. Time} &
\colhead{Wavelength or Filters} &
\colhead{Resolution}  &
\colhead{Reference} 
}
\startdata 
Spectroscopy  &    &   &        &       &                               \\
19-20 Sep. 2006 & WIYN/Hydra  & 180m & 3970--5030{\AA}    & 1.5{\AA} & \citet{Massey07} \\
21-22 Sep 2006  & WIYN/Hydra  & 90m &  4550--7400{\AA}   & 3.4{\AA}  & \citet{Massey07} \\
10 Oct. 2010  & MMT/Hectospec & 120m  &  3600--6050{\AA}  & 1.9-2.2{\AA} &  Humphreys et al. 2013,2014)\\
09 Oct. 2010  & MMT/Hectospec & 90m  &   5500--8000{\AA}  & 1.9-2.2{\AA} & Humphreys et al. 2013,2014)\\ 
25 Sep. 2013  & MMT/Hectospec & 120m  &  3600--6050{\AA}   & 1.9-2.2{\AA}   &  This paper \\
26 Sep. 2013  & MMT/Hectospec &  90m  &  5500--8000{\AA}  & 1.9-2.2{\AA}   & This paper \\
              &               &       &       &               &            \\
Photometry    &               &       &       &                &            \\ 
1999 - 2003   &  WFC/INT      &  \nodata  &  BV    & \nodata  & \citet{Vilardell}\\ 
03 -06 Oct. 2000 & Mosaic CCD/4m & \nodata & UBVRI  & \nodata   & \citet{Massey06}\\ 
13 Aug. 2010  & HST/ACS        & \nodata &  BI   & \nodata  & This paper \\
Oct.-Nov. 2011 & BTA 6-m.      & \nodata &  BVR    & \nodata  & \citet{Shol}\\ 
11 Jan. 2012  & HST/ACS        & \nodata &  BI  & \nodata   & This paper \\
02 Sep. 2013     &CCD/Barber Obs. & \nodata &  VR  & \nodata  & This paper \\
28 Oct. 2013     &CCD/Barber Obs. & \nodata &  VR  & \nodata  & This paper \\
21 Oct. 2014     &CCD/Barber Obs. & \nodata &  BVR   & \nodata  & This paper \\ 
21 Nov  2014     &CCD/Barber Obs. & \nodata &  BV   & \nodata & This paper \\ 
15 Jan  2015     &CCD/Barber Obs. & \nodata &  BVR & \nodata & This paper \\
\enddata
\end{deluxetable} 

\begin{deluxetable}{lllll}
\tabletypesize{\footnotesize}
\tablenum{2}
\tablecaption{Multi-color Photometry 2000 - 2015}
\tablehead{
\colhead{Date}  &
\colhead{B mag} &
\colhead{V mag} & 
\colhead{R mag} &
\colhead{I mag}
}
\startdata
1999 - 2003 &  17.11 & 17.08 & \nodata & \nodata \\ 
03 - 06 Oct. 2000 &  17.66 & 17.16 & 17.48 &  17.35 \\
13 Aug. 2010 &  17.31  & \nodata & \nodata  &  16.61 \\
Oct.-Nov. 2011 & 16.82 & 16.37 & 16.08 & \nodata \\
11 Jan. 2012  &  16.81  &  \nodata & \nodata &  15.93  \\
02 Sep. 2013  &  \nodata & 16.23 & 15.93 & \nodata \\
28 Oct. 2013  &  \nodata &  16.27 & 15.99 & \nodata \\
21 Oct. 2014  &  16.66   &  16.46 &  16.17 & \nodata \\
21 Nov. 2014  &  16.68   &  16.45 &  \nodata  & \nodata \\
15 Jan. 2015  &  16.76   &  16.51 &  16.27   & \nodata \\ 
\enddata
\end{deluxetable} 

\begin{deluxetable}{lll}
\tabletypesize{\footnotesize}
\tablenum{3}
\tablecaption{Outflow Velocities in J004526.62+415006.3}
\tablehead{  
\colhead{Date}  &
\colhead{Velocity km s$^{-1}$}  &
\colhead{Lines} 
}
\startdata
Sep. 2006  &  -101 $\pm$ 3  & 3 Hydrogen lines \\              
Oct.2010  &  -149 $\pm$ 6  & 4 Hydrogen lines \\
          &  -138 $\pm$ 3  & 3 Fe II lines \\
Sep.2013  & -115  &  H$\alpha$ only \\ 
\enddata
\end{deluxetable}

\end{document}